\documentclass[11pt]{article}
\usepackage{hyperref}
\usepackage{graphicx}
\pdfoutput=1
\begin{document}
\title{Bursting of rigid bubbles}
\author{Pauline C. Petit and Anne-Laure Biance\\
\\\vspace{6pt} Institut Lumi\`ere Mati\`ere, UMR5306, Universit\'e Lyon 1\\ Domaine Scientifique de La Doua, 10 rue Ada Byron\\
69622 Villeurbanne CEDEX, France}
\maketitle
%% The abstract (in this file, and that submitted as text to arXiv) should
%include the exact phrase
%% "fluid dynamics video" or "fluid dynamics videos"
\begin{abstract}
We propose here a fluid dynamics video relating the bursting of soap rigid films. 
\end{abstract}
% main text

\graphicspath{{}}

\section{Introduction}

%The {\em hyperref} package is used to make links to the videos.
%% The format is: \href{URL of video}{name that will appear in the text}

\begin{figure}[t]
\centering
\includegraphics[width=.9\columnwidth]{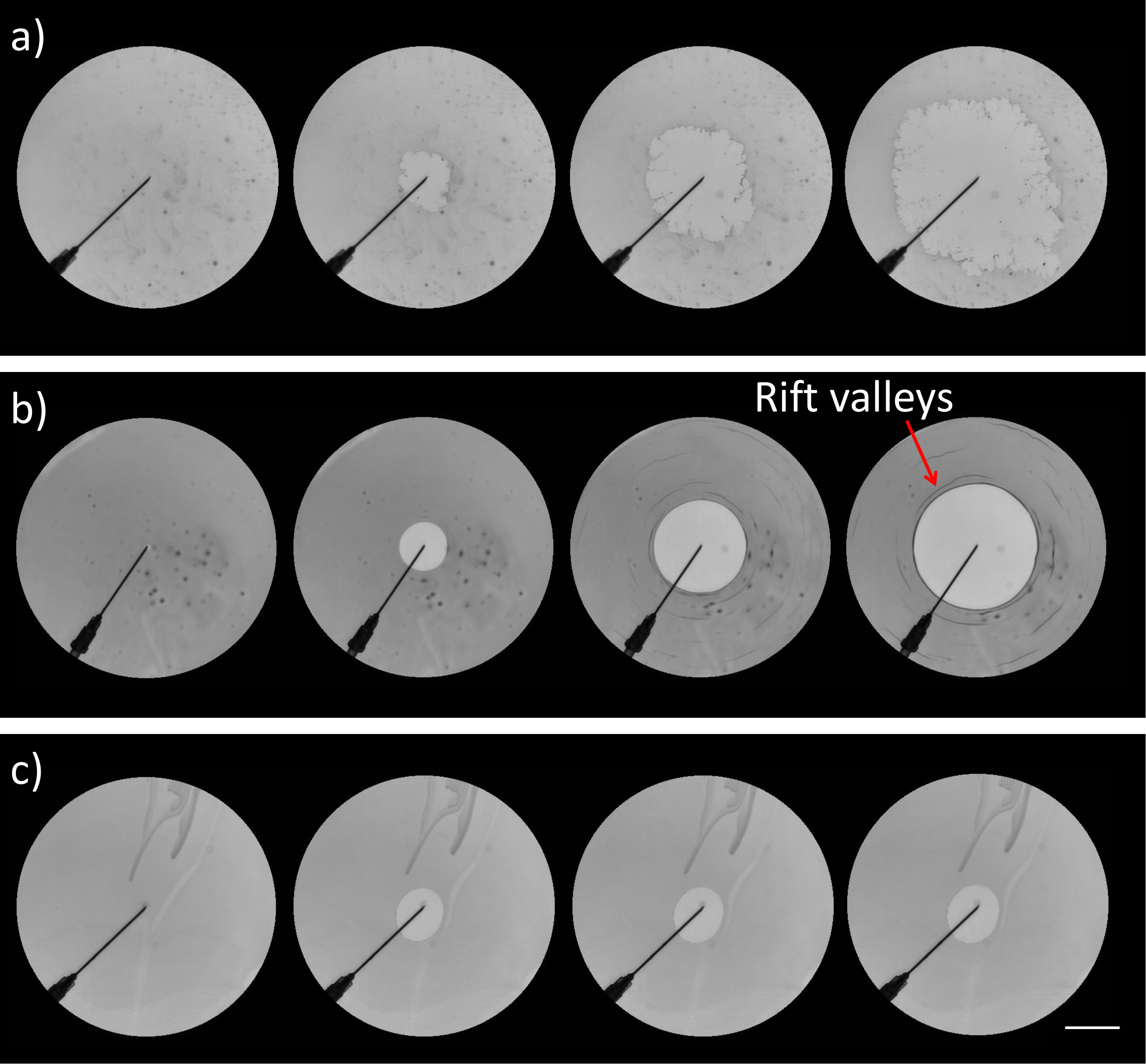}
\caption{Image sequences of film bursting obtained with mobile interfaces (a) and rigid interfaces (b and c), for thicknesses of 1 $\mu$m (a), 10 $\mu$m (b) and 3 $\mu$m (c). The time interval between the images is 1.1 ms (a) or 3.7 ms (b and c). The scale bar represents 20 mm.}
\label{sequence}
\end{figure}

The dynamics of destabilization of soap bubbles is an old well studied issue. Lucien Bull (1904) with Etienne-Jules Marey chronophotographer made the first images of soap bubble bursts. The use of high speed camera suggests that the dynamics of the hole opening is governed by inertia. Taylor \cite{Taylor59} proposed that hole opening dynamic result from a force balance of a rim at the edge of the hole:
\begin{equation}
\frac{d}{dt} (mV)= 2 \gamma
\end{equation}
where $m$ is the mass of the rim, varying with time as the rim swallows some part of the film, and $V$ its velocity. The resolution of this equation shows a constant velocity of the rim:
\begin{equation}
V=\sqrt{\frac{\phi \gamma }{\rho h}}
\label{eqiner}
\end{equation}
where $\phi$ has been demonstrated to be equal to 2 one year later by Culick \cite{Culick60}. These results are in good agreement with stationary experiments performed on liquid sheet \cite{Taylor59} and has been extensively investigated by McEntee and Mysels in the case of soap film \cite{McEntee1969}. During the rupture, the rim has been shown to destabilize into satellite droplets with a controlled size and with a well-defined wavelength \cite{Lhuissier2009}, but without any influence on soap film bursting dynamics.

\begin{figure}[bthp]
\centering
\includegraphics[width=.7\columnwidth]{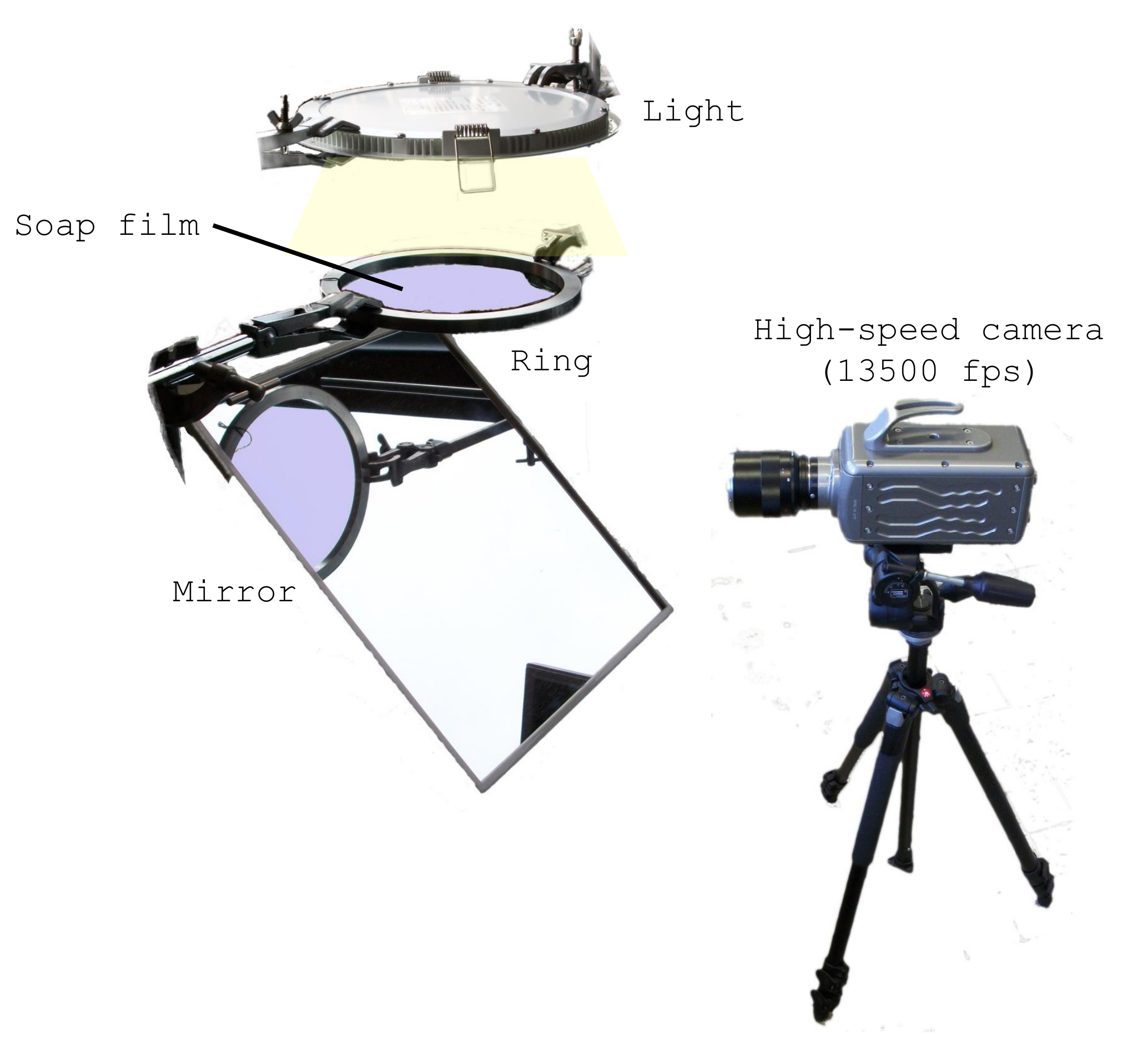}
\caption{Experimental setup to capture movies of soap film bursts.}
\label{manip}
\end{figure}

A deviation from this law is observed for $rigid$ films (SDS+LOH) \cite{McEntee1969}, but no systematic studies have provided quantitative results for this effect. We investigate it by using an insoluble surfactant mixture \cite{Golemanov:2008hc}, generating the so-called $rigid$ interfaces. The experimental set-up consists of a metallic frame, dipped in a surfactant solution (mix of SLES, 0.33 wt\%, CAPB, 0.17 wt\% and Myristic acid, 0.02\%) with added dye (Brilliant Black BN, 0.5 wt\%) and 10\% of glycerol. The metallic frame and subsequent generated soap film is withdrawn from the bath with a motor at a constant velocity. Motor velocity variations allow to tune film thickness, which is measured by image gray level analysis and adequate calibration through Beer-Lambert light absorption of the dye (Figure \ref{manip}). The measurement method has been verified by experiments with a solution of SDS (mobile interfaces) which give a good agreement with the Taylor-Culick's law (Figure \ref{sequence}a).

For rigid interfaces, depending on film thickness, different observations can be performed. At large and intermediate thicknesses (between 3 and 20 $\mu$m), the hole radius increases with time but endorses a deceleration. The initial velocity is shown to be smaller than Taylor-Culick expected one. Images demonstrate the formation of rift valleys within the films, on which discontinuities of the velocity in the liquid are observed (Figure \ref{sequence}b). At smaller thicknesses (below 3 $\mu$m), the hole opening dynamics can be divided into three steps (Figure \ref{sequence}c). A first fast opening (but slower than Taylor-Culick's expectations), is followed by a second step where the dynamic is almost stopped,  and which is associated with the formation of rift valleys. Then the dynamics increases again. The rift valleys can join together making a pattern similar to rupture propagation in solid materials.

\bibliographystyle{plain}
\bibliography{gallery}

\begin{thebibliography}{1}

\bibitem{Culick60}
F.~E.~C. Culick.
\newblock Comments on a ruptured soap film.
\newblock {\em {Journal of applied physics}}, {31}({6}):{1128--1129}, {1960}.

\bibitem{Golemanov:2008hc}
K.~Golemanov, N.~D. Denkov, S.~Tcholakova, M.~Vethamuthu, and A.~Lips.
\newblock Surfactant mixtures for control of bubble surface mobility in foam
  studies.
\newblock {\em Langmuir}, 24(18):9956--9961, Sep 2008.

\bibitem{Lhuissier2009}
H.~Lhuissier and E.~Villermaux.
\newblock Soap films burst like flapping flags.
\newblock {\em Phys. Rev. Lett.}, 103:054501, Jul 2009.

\bibitem{McEntee1969}
W.~R. McEntee and K.~J. Mysels.
\newblock Bursting of soap films. i. an experimental study.
\newblock {\em The Journal of Physical Chemistry}, 73(9):3018--3028, 1969.

\bibitem{Taylor59}
G.I. Taylor.
\newblock {The dynamics of thin sheets of fluid .3. disintegration of fluid
  sheets}.
\newblock {\em {Proc. Royal Soc. A}}, {253}:{313}, {1959}.

\end{thebibliography}

%Two sample videos are

\end{document}